\documentclass[twocolumn,pre,showpacs,amsfonts,amsmath,assymb,eufrak
,longbibliography]{revtex4-2}
\usepackage{epsfig}
\usepackage{color}
\usepackage{bm}
\usepackage[
colorlinks=true, 
pdfstartview=FitV, 
linkcolor=blue, 
citecolor=magenta, 
urlcolor=blue, 
bookmarks=true,
bookmarksnumbered=true,
pdftitle={},
pdfauthor={}
]{hyperref}
\usepackage{epsfig,amsopn}
\usepackage{graphicx}

\usepackage{amssymb}
\usepackage{amsmath}
\usepackage{color}
\usepackage{subfigure}
\usepackage{textcomp}
\usepackage[utf8]{inputenc}
\definecolor{pink}{rgb}{1,0,0.6}

\newcommand\nn{\nonumber}

\usepackage{grffile}
\graphicspath{{./Figures/}{./Hydro/}{./SpaceAverage/}}
\usepackage{textcomp}

\newcommand{\be}{\begin{equation}}
\newcommand{\ee}{\end{equation}}
\newcommand{\bea}{\begin{eqnarray}}
\newcommand{\eea}{\end{eqnarray}}

\begin{document}

\newcommand{\titlename}{Fluctuating Hydrodynamics Describes Transport in Cellular Aggregates}

\title{\titlename}

\author{Subhadip Chakraborti$^{1,2}$ and Vasily Zaburdaev$^{1,2}$}

\affiliation{$^1$Department of Biology, Friedrich-Alexander-Universit\"{a}t Erlangen-N\"{u}rnberg, 91058 Erlangen, Germany}%
\affiliation{$^2$Max Planck Zentrum für Physik und Medizin, 91058 Erlangen, Germany}

\date{\today}
\begin{abstract} 
Biological functionality of cellular aggregates is largely influenced by the activity and displacements of individual constituent cells. From a theoretical perspective this activity can be characterized by hydrodynamic transport coefficients of diffusivity and conductivity. Motivated by the clustering dynamics of bacterial microcolonies we propose a model of active multicellular aggregates and use recently developed macroscopic fluctuation theory to derive a fluctuating hydrodynamics for this model system. Both semi-analytic theory and microscopic simulations show that the hydrodynamic transport coefficients are affected by non-equilibrium microscopic parameters and significantly decrease inside of the clusters. We further find that the Einstein relation connecting the transport coefficients and fluctuations breaks down in the parameter regime where the detailed balance is not satisfied. This study offers valuable tools for experimental investigation of hydrodynamic transport in other systems of cellular aggregates such as tumor spheroids and organoids.
\end{abstract}
\maketitle

\section{Introduction}
\label{intro}
Multicellular aggregates consisting of a large number of interacting individual cells are ubiquitous and span a wide spectrum of biological systems, including bacterial colonies \cite{Gutnick1998, Zaburdaev2015, Zaburdaev2017}, tumor spheroids \cite{Wolfgang_Review1987, Friedl_NatureReview2003, Fabry_PLoS2012, Fabry_eLife2020}, and stem cell organoids \cite{Karow_Nature2018, Karow_2021, Lancaster_Nature2013, Pasca_Nature2018}. The complex biological behaviors observed within these aggregates typically emerge as a collective outcome of individual cellular movements, referred to as {\it activity}, in which cells convert chemical energy into a mechanical one. This activity maintains the system in a state out of equilibrium, giving rise to diverse collective phenomena such as self-assembly and pattern formation, that would otherwise be unattainable under equilibrium conditions \cite{Marchetti_Review2013, TonerTu_PRL1995, Ramaswamy_Review2010, Cates_Review2015, Marchetti_PRL2012, Golestanian_PRL2007, Raphael_Cell2018, Rafael_PRE2022, Manning_JRS2013, Manning_NP2015}.

Complex temporal history of microscopic active fluctuations experienced by the cell within the aggregate plays a pivotal role in biological cell function, such as phenotype differentiation in bacterial colonies \cite{ponisch2018pili}, cell fate specification in organoids \cite{tortorella2021role} or developing antibiotic resistance by controlling transport within bacterial colonies \cite{Maier2021}. This signifies the importance of the fluctuating hydrodynamics approach as a framework connecting the large-scale spatio-temporal transport to microscopic active interactions via macroscopic fluctuations. Notably, unlike in equilibrium \cite{Green_1954,Kubo_1966}, there is presently no comprehensive theoretical understanding regarding the potential connections between transport coefficients and macroscopic fluctuations in active systems with a number of important works starting to emerge in phenomenology \cite{Ramaswamy_PRL2002,Cates_PRL2013, Cates_PRX2017,Golestanian_PRE2019}, using gradient expansion method \cite{Chate_PRL2012, Tailleur_PRL2012, Speck_EPL2013, Baskaran_JStatMech2017}, for systems with very few particles \cite{Slowman_PRL2016, Malakar_2018, Majumdar_PRL2020, Mohanty_SPP2023} and with many particles \cite{Tailleur_PRL2013, Tailleur_PRL2018, Chakraborty_PRE2020, Rahul_PRE2020, Agranov_JSM2021, Agranov_SPP2023, Agranov_JSM2022, Dandekar_JSM2023, Tanmoy_PRE2024, Tanmoy2023}.

In this paper, motivated by the system of interacting, colony forming {\it Neisseria gonorrhoeae} bacteria we develop fluctuating hydrodynamics theory of dense cellular aggregates. These and many other bacteria exhibit stochastic but active movements on a surface and form microcolonies via active contractile intercellular forces. By formulating one dimensional stochastic model of this system we show that the interplay between cell-cell and cell-substrate interaction leads to a clear clustering transition. We derive fluctuating hydrodynamic equation for particle (cell) density field using a recently developed macroscopic fluctuation theory (MFT) \cite{Bertini_RMP2015}. It enables us to obtain two density dependent transport coefficients -- bulk diffusivity and conductivity, quantities that can be directly assessed experimentally. Our semi-analytical analyses supported by microscopic simulations demonstrate a notable reduction in hydrodynamic transport coefficients due to influence of the non-equilibrium microscopic parameters within a clustered regime. However, despite of being a \textit{gradient type} model working within the framework of MFT, the Einstein relation, which links transport and fluctuation properties, is violated in the same regime of clustering. Microscopically, this can be attributed to the departure from detailed balance resulting from active cell-cell interactions. However, a macroscopic explanation for this unusual phenomenon remains an area for potential future research. 
This study presents practical methodology of obtaining hydrodynamic transport in active systems i) driven by active dipole force between particles ii) with short (exponential) interaction range and iii) stochastic life time. This set of properties is rather generic and we believe is pertinent to broad range of cellular aggregates, including but not limited to bacterial microlonies, tumor spheroids, and organoids.

The paper is organized as follows. In Sec.~\ref{sec:model} we define the one-dimensional stochastic model of aggregation and  formulate it as an exclusion model and its mapping to an unbounded model. In Sec.~\ref{sec:hydro} we derive a fluctuating hydrodynamic description starting from microscopic picture and obtain the transport coefficients semi-analytically. Sec.~\ref{sec:result} shows the numerical results and comparison with the semi-analytic theory. We conclude in Sec.~\ref{conclusion} with the description of the main results and a discussion of the remaining open questions.

\begin{figure}
\begin{center}
\includegraphics[width=8.5cm,angle=0]{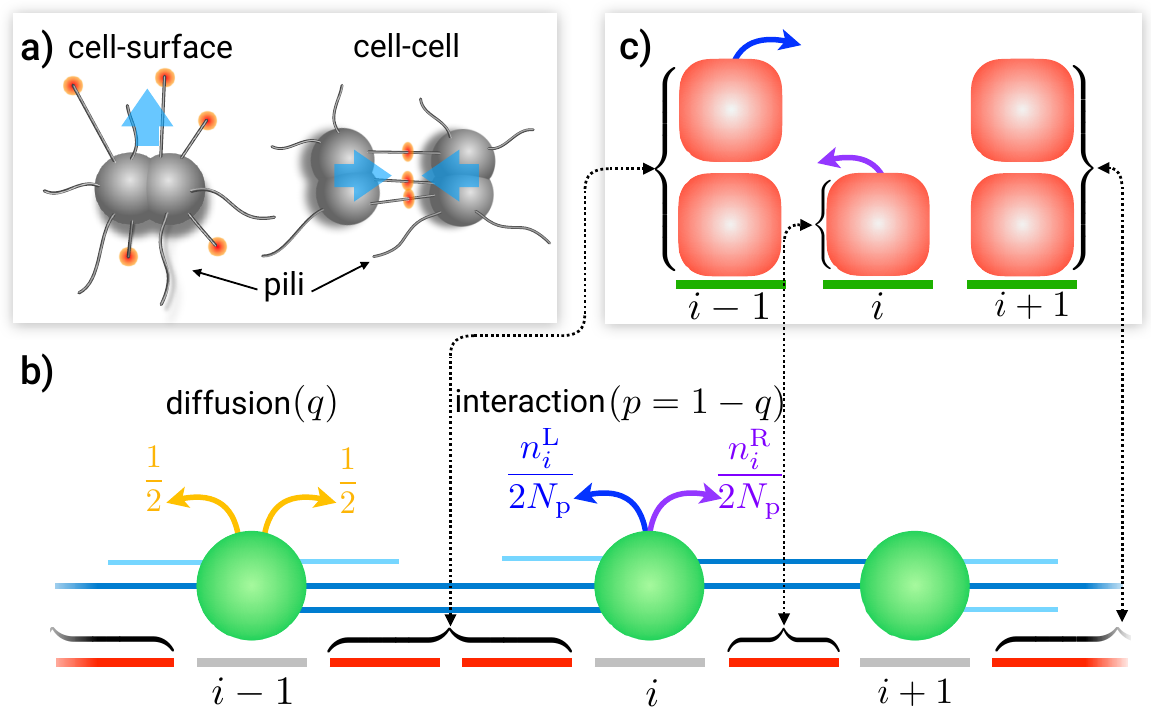}
\caption{Panel (a): Schematic diagram of microscopic forces generated by \textit{N. gonorrhoeae} bacteria cells.  Diplococcus shaped bacteria use long retractile filaments called type IV pili to attach to a surface (left) and to each other (right). Pili attached to the surface (indicated with red circles) retract and generate pulling force moving cell forward (blue arrow). When pili of neighboring cells are attached, their contraction results in a force dipole pulling cells together (blue arrows). Panel (b): Our exclusion model (EM) of \textit{N. gonorrhoeae} bacteria clustering on 1D periodic lattice. Not bound (free) pili are shown in light blue while bound pili-pairs indicated by dark blue. Hopping rates due to pili-substrate (occurring with stochastic weight $q$) and bound pili-pair interactions (occurring with stochastic weight $p=1-q$) are shown respectively by yellow and blue-purple arrows. For simplicity we assume that pili-surface interactions lead to a simple random diffusive hopping. Pili-pili interactions bias the hopping rate proportional to the number of pili bound on each side of a particle ($n^L$ and $n^R$) where $N_p$ is constant total number of pili per particle. Panel (c): Mapping of EM to an unbounded mass transfer model (UMTM). Number of empty lattice sites to the right of the particle in the EM model  corresponds to the number of particles in the UMTP model. Note that the movement of mass in UMTM is in opposite direction to that of EM.}
\label{model}
\end{center}
\end{figure}

\section{The model}
\label{sec:model}

{\it N. gonorrhoeae} and many other bacteria species, employ thin, flexible filaments called type IV pili for interactions within their environment and among each other \cite{Tainer2004, Mattick2002} [see Fig. \ref{model}(a)]. Each {\it N. gonorrhoeae} bacterium possesses multiple $(\sim 10-20)$ pili uniformly distributed  across its surface \cite{Zaburdaev2017, Maier_PRL2010}. The growth of pili \cite{SheetzMP_Nature2000,Zaburdaev2017,Berg_PNAS2001,Maier2019} results in random pili lengths following an exponential distribution with an average length of $l_0\sim 1-2 ~\mu$m comparable to the cell size of $\sim 1 ~\mu$m. Attachment of pili to a substrate and their retraction generates the force powering twitching cell motility \cite{Mattick2002,Ellison,Henrichsen}. The stochastic binding between two individual pili of neighboring cells and their retraction generates an attractive force dipole ($\sim 180$~pN) and mediates formation of cell clusters. The stochastic detachment of pili impacts the force balance by altering both the number of bound pili-pairs and of pili bound to a substrate. 

We now construct a one dimensional model recapitulating the above described dynamics. Individual bacteria are represented by $N$ non-overlapping particles that can move on a periodic lattice of size $L$ [see Fig.~\ref{model}(b)]. The $N_p$ number of pili associated with each particle can be created to the left and to the right of the particle with exponentially distributed length $l$ \cite{Maier_PRL2010} and the mean $l_0$. The pili turnover process is incorporated by attaching a `clock' to every pilus with a random lifetime $T$ which is exponentially distributed with a mean lifetime $T_0$. As the lifetime elapses, the pilus (free or bound) is destroyed and a new pilus is created randomly to the left or right of the same particle thus keeping the total number of pili $N_p$ per particle conserved. The stochastic weights associated with particle moves (to the neighboring site) due to pili-substrate and pili-pili interactions are respectively $q$ and $p$, with $p+q=1$. Pili dynamics affects both, cell motion on a substrate and cell clustering \cite{Zaburdaev_PRE2015, Zaburdaev_PRL2021}. 
The pili-driven cell-substrate interaction lead to an effective persistent motion of interacting run-and-tumble-like particles which can result in clustering as was previously explored in Ref.~\cite{Rahul_PRE2020} in the context of fluctuating hydrodynamics. While pili-driven surface motility is relatively well understood, dynamics of clusters emerging due to active, pili-driven cell-cell interactions is virtually unexplored.
Thus, we simplify the effect of pili-substrate interaction to a symmetric hopping by one unit lattice with equal probability $1/2$. The particle dynamics due to pili-pili interaction is more involved. If the total length of two pili pointing towards each other from two adjacent particles is greater or equal to the distance between the particles then those pili will form a bound pair. A particle will advance one unit lattice spacing toward left or right according to random weight $n^L/(2N_p)$ and $n^R/(2N_p)$ respectively where $n^L$ and $n^R$ are the number of bound pairs in the left and right directions respectively. Only next neighboring particles can interact irrespective of their distance in between.  
\begin{figure}
\begin{center}
\includegraphics[width=8.5cm,angle=0]{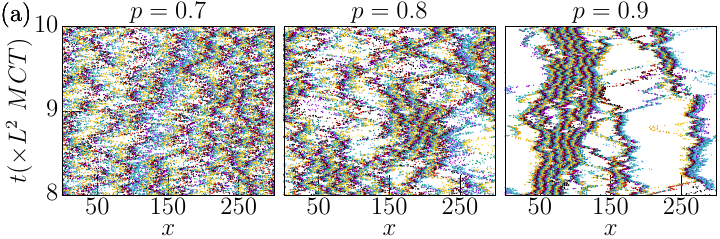}
\includegraphics[width=4.27cm,angle=0]{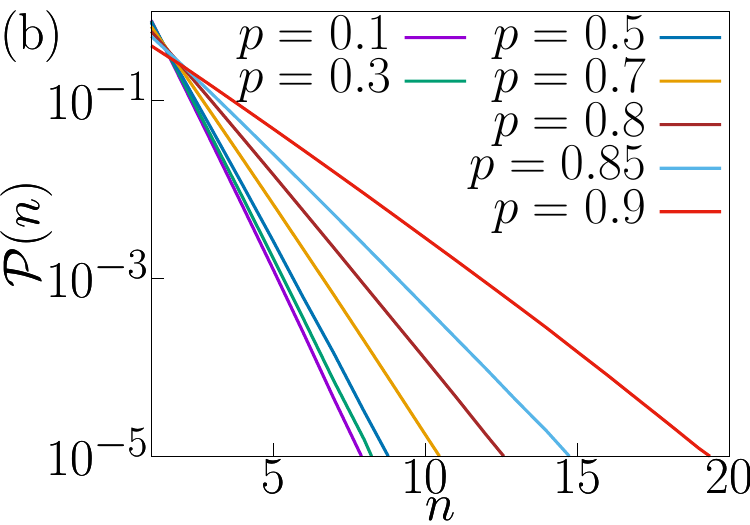}
\includegraphics[width=4.27cm,angle=0]{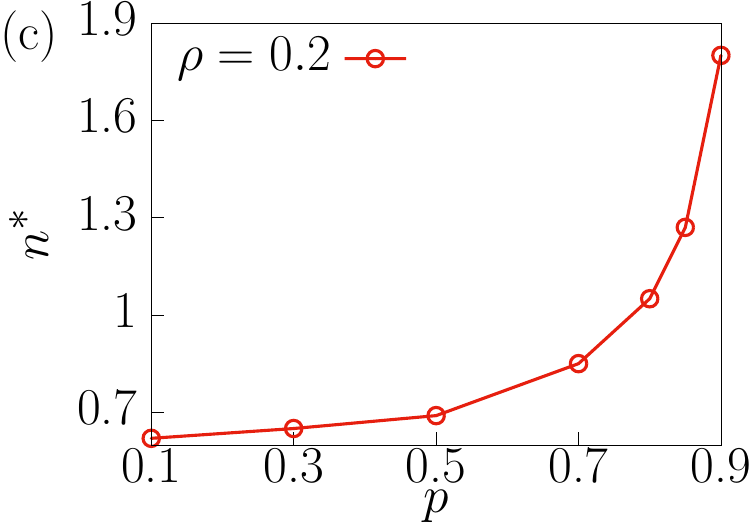}
\caption{Panel (a): Trajectories of $N=60$ particles in a 1D box of size $L=300$. We see clustering transition by varying cell-cell interaction parameter $p=0.7$, $0.8$ and $0.9$ while keeping other parameters fixed at $T_0=10$, $l_0=2$, and $N_p=10$. Panel (b): Probability distribution ${\cal P}(n)$ of cluster size $n$. The distribution is exponential with the tail becoming `heavier' with increment of $p$. Panel (c): Typical cluster size $n^*$ is obtained from a phenomenological fit of an exponential function ${\cal P}(n) \sim \exp{(-n/n^*)}$ to numerically obtained distributions from panel (b) (lines connecting dots serve to guide the eye).  Its increase with increasing $p$ indicates a clustering transition. Other parameters are same as in previous panels.}
\label{trajectories}
\end{center}
\end{figure}
The model involves several experimentally motivated parameters such as density $\rho=N/L$, stochastic weight of cell-cell interaction $p$, mean pili-lifetime $T_0$, number of pili per particle $N_p$ and mean pili length $l_0$ which all affect collective behaviors. In this work, we primarily focus on the effect of $\rho$ and the cell-cell interaction parameter $p$, but will also briefly discuss the effects of other parameters. 

We first look into the clustering phenomena by observing the trajectories of the particles from stochastic simulations. In panel (a) of Fig.~\ref{trajectories}, we see a clustering transition from a homogeneous state to a clustered state as we increase the pili-pili interaction rate $p$ while keeping all other parameters fixed ($\rho=N/L=0.2$, $T_0=10$ measured in Monte Carlo time step (MCT) where one unit of MCT gives on average a chance to each particle to jump, $N_p=10$, and $l_0=2$ measured in units of lattice spacing $\ell$). By defining a cluster as $n$ consecutive occupied sites we numerically calculate the cluster-size distributions ${\cal P}(n)$, see panel (b) of Fig.~\ref{trajectories}. We always observe an exponential distribution with a `heavier' tail as we increase $p$ indicative of clustering. Fitting the distributions with ${\cal P}(n) \sim \exp{(-n/n^*)}$ quantifies the typical cluster size $n^*$. The plot of the typical cluster size $n^*$ with $p$ in the panel (c) in Fig.~\ref{trajectories} also supports our finding with continuous transition towards clustering.  Interestingly, if we define clusters more loosely as a set of particles that are bound by pili pairs, the cluster-size distribution changes a shape and attains a ``hump'' at the tail of the distribution (see Appendix D). This is similar to the experimentally observed cluster-size distribution in {\em N. gonorrhoeae} colonies \cite{Zaburdaev2015}.

Next we aim to investigate how the macroscopic transport properties such as diffusivity and conductivity behave during this transition by deriving a fluctuating hydrodynamics based on the framework of MFT. To this aim we first show how to map our exclusion model to the unbounded mass model.

\subsection{Mapping to unbounded model}

Analytical treatment of exclusion models with non-trivial interactions is often very challenging, particularly when one faces calculations of correlations emerging from the microscopic interactions. Here we use an elegant technique \cite{Evans_JPA2005} which simplifies the calculation of correlations by an exact mapping of the exclusion model (EM) to an unbounded mass transfer model (UMTM) and allows us to move forward with MFT to derive fluctuating hydrodynamics. The rules of mapping are as follows [see Fig.~\ref{model}(c)]. The bottom panel (b) is our original EM where occupancy of a site by a particle (green circle) can be either zero or unity. In the top right panel (c), we construct a new 1D ring (UMTM) consisting of $\tilde{L} = N$ sites where number of masses (red squares) at any site is unbounded. For every $i^{\mbox{th}}$ (green) particle on the original EM, we construct a (green) lattice site labeled by $i$ in the UMTM. For $m_i$ vacant lattice sites in between $i^{\mbox{th}}$ and $(i+1)^{\mbox{th}}$ particles in the EM we put (red) mass $m_i$ at $i^{\mbox{th}}$ (green) site of the UMTM, so that the total mass in the UMTM is $\tilde{N} = \sum_i m_i=L-N$. Thus, the mass density $\tilde{\rho}$ in the UMTM is linked to EM density $\rho$ via: $\tilde{\rho} = \tilde{N} / \tilde{L} = 1/\rho -1$. Note that the movement of mass in UMTM is in the opposite direction of the particle movement in the EM. We next use this mapping to derive hydrodynamics and to calculate transport coefficients in UMTM and then use a reverse mapping to recover results in the original EM setting.

\section{Derivation of hydrodynamics} 
\label{sec:hydro}

We first put forward the master equation by considering all possible ways of transferring a mass from site $i$ to its neighboring site $j$ with jump weights $c_{i j}$. Thus, for UMTM, the continuous time evolution of mass $m_i(t)$ at site $i$ and at time $t$ in an infinitesimal time interval $dt$ is given by 
\begin{eqnarray}
m_i(t+dt) =
\left\{
\begin{array}{ll}
m_i(t) - 1            & {\rm prob.}~ c_{i,i-1} dt/2, \\
m_i(t) - 1            & {\rm prob.}~ c_{i,i+1} dt/2, \\
m_i(t) + 1            & {\rm prob.}~ c_{i-1,i} dt/2, \\
m_i(t) + 1            & {\rm prob.}~ c_{i+1,i} dt/2, \\
m_i(t)                & {\rm prob.}~ 1-\Sigma dt,
\end{array}
\right.
\label{unbounded-unbiased}
\end{eqnarray}
with $\Sigma=\left(c_{i,i-1}+c_{i,i+1}+c_{i-1,i}+c_{i+1,i}\right)/2$. The explicit expressions of the jump weights are: $c_{i,i-1}=qa_i+pa_in^R_i/N_p$, $c_{i,i+1}=qa_i+pa_in^L_{i+1}/N_p$, $c_{i-1,i}=qa_{i-1}+pa_{i-1}n^L_i/N_p$ and $c_{i+1,i}=qa_{i+1}+pa_{i+1}n^R_{i+1}/N_p$ with $a_i=1-\delta_{m_i,0}$ is the probability for a site $i$ to be occupied. Note that in EM, any two neighboring particles share the same number of bound pairs between them implying $n^R_i=n^L_{i+1}$ which gives our UMTM system a structure of zero-range-process where the jump weights $c_{i,j}=qa_i+pa_in^R_i/N_p$ depend only on the initial site $i$. Here, all lattice indices are associated with a length-scale which is the lattice constant $\ell$ which we set to unity for simplicity and the time $t$ is measured in MCT. In supplemental material \cite{supplement}, we show the master equation structure for the local mass density $\tilde{\rho}_i(t)=\langle m_i(t) \rangle$ given the jump rates in Eq.~\eqref{unbounded-unbiased} and arrive at:
\begin{eqnarray}
\partial_t \tilde{\rho}_i(t) = \frac{1}{2}\left( G_{i-1}+G_{i+1}-2G_i \right), G_i=q A_i + \frac{p}{N_p}U_i
\label{gradient1}
\end{eqnarray}
where a local observable $A_i=\langle a_i \rangle$ is the occupation probability in UMTM and $U_i=\langle a_i n^R_i \rangle$ is the average number of bound pili-pairs on the right side of site $i$ provided that the site is occupied. $A_i$ and $U_i$ are the only two important observables that are required to derive the hydrodynamics. Note that the local current in Eq.~\eqref{gradient1} has a discrete gradient structure of local observable $G_i$ which makes our system a {\it gradient type} \cite{Krapivsky_PRE2014}. This gradient condition allows us to write down the time-evolution equation for the coarse-grained density field $\tilde{\rho}(x,\tau)$ in the diffusive scaling limit $i \rightarrow x = i/\tilde{L}$ and $t \rightarrow \tau = t/\tilde{L}^2$ after neglecting higher order terms ${\cal O}(1/\tilde{L}^3)$ for large system size $\tilde{L}$ as \cite{supplement}:
\begin{equation}
\frac{\partial \tilde{\rho}(x,\tau)}{\partial \tau} =  - \frac{\partial}{\partial x} \left( -D(\tilde{\rho}) \frac{\partial \tilde{\rho}}{\partial x} \right).
\label{diffusion}
\end{equation}
Here the diffusive current field $J_D(\tilde{\rho})=-D(\tilde{\rho}) \partial_x \rho$ involves the transport coefficient of bulk-diffusivity $D(\tilde{\rho})=(\partial G(\tilde{\rho})/\partial \tilde{\rho})/2$ which in general also depends on other parameters $p$, $T_0$, $N_p$ and $l_0$. It is important to note that the bulk-diffusivity $D(\rho)$ differs from the self-diffusivity of a tagged particle \cite{Alexander_PRB1978, Krapivsky_JSP2015, Krapivsky_PRL2014}. Bulk diffusivity relates the mean (hydrodynamic) current of particles to their density gradient, whereas self-diffusivity characterizes dependence of the mean-squared displacement (MSD) of a tagged particle as a function of time. In a many-body system of interacting particles these diffusivities do not necessarily coincide.
Eq.~\eqref{diffusion} is a deterministic hydrodynamic equation that describes `typical' coarse-grained trajectories of the system. The behavior of the `atypical' trajectories of the hydrodynamic quantities such as coarse-grained density or coarse-grained current can be captured from the fluctuations of the respective hydrodynamic quantities and those macroscopic fluctuations can bring forward new transport coefficients such as conductivity. This essentially requires the derivation of fluctuating hydrodynamics which we do next.

One useful tool to derive fluctuating hydrodynamics is the recently developed macroscopic fluctuation theory \cite{Bertini_PRL2001, Bertini_RMP2015, Derrida_JSM2007} applicable to stochastic diffusive systems having Markov property where the total number of particles is conserved. In last few decades, MFT has served to develop fluctuating hydrodynamics in several boundary driven and other lattice gas systems \cite{Bertini_RMP2015, Derrida_JSM2007, Derrida_PRL2004, Derrida_JSP2009, Derrida_JSM2009, Krapivsky_PRE2012, Krapivsky_PRL2014}, non-equilibrium mass transport models \cite{Das_PRE2017, Chakraborti_PRE2021} as well as in some inherently out-of-equilibrium models of active matter \cite{Chakraborty_PRE2020, Rahul_PRE2020, Agranov_SPP2023, Dandekar_JSM2023, Tanmoy2023}. MFT suggests that for a process of gradient type, one would expect a fluctuating hydrodynamics equation to be written in the form of a continuity equation \cite{Bertini_PRL2001, Bertini_RMP2015, Derrida_JSM2007}:
\begin{equation}
\frac{\partial \tilde{\rho}(x,\tau)}{\partial \tau} =  - \frac{\partial}{\partial x} \left( -D(\tilde{\rho}) \frac{\partial \tilde{\rho}}{\partial x} +\sqrt{\chi(\tilde{\rho})} \eta(x,\tau) \right),
\label{fluc-hydro}
\end{equation}
where conductivity $\chi(\tilde{\rho})$ is another density dependent transport coefficient of interest for us and $\eta(x,\tau)$ is a coarse-grained noise term -- white but not necessarily Gaussian in general.
This conductivity describes the hydrodynamic drift current as a response to a force -- internal random or external deterministic force. The internal cell-generated forces are random forces that give rise to a fluctuating current with a strength equal to the square root of conductivity [as seen in Eq.~\eqref{fluc-hydro}]. This random current, however, vanishes when averaged over time or realizations. A straightforward way to determine $\chi(\tilde{\rho})$, according to Eq.~\eqref{fluc-hydro}, is to calculate the variance of the coarse-grained current by evaluating the probability of rare events, when the path history of the coarse-grained variables deviates from the typical evolution governed by the typical current $J_D(\tilde{\rho})$. These rare events occur with an exponentially small probability in the system of a large size, giving rise to a large deviation principle. This is why it is always more difficult than calculating the mean current. However, MFT suggests an alternative and easier way to obtain $\chi(\tilde{\rho})$ by biasing the system with a small external force $F$ such that the non-stationary large deviation current and density fields of the original unbiased system become the typical current and density fields for the biased one. In a biological system, a relevant external force could be a gravitational force when a substrate on which cells move is inclined relative to the horizontal orientation. It is hypothesized that, such a small bias does not generate an additional non-equilibrium contribution to the system and thus satisfies a local detailed balance condition without affecting the inherent non-equilibrium nature of the system. By applying a small constant biasing force field $F$, for example towards the right direction, the mass transfer weights in the biased system $c^F_{i j}$ are obtained by exponentially weighting the original rates $c_{i j}$ of Eq.~\eqref{unbounded-unbiased} with a proper normalisation as:
\begin{eqnarray}
c^F_{i j} =\frac{ 2 c_{i j} \exp\left[ \frac{1}{2} \hat{m}_{i j}F(j-i) \ell \right]}{\exp\left[ \frac{1}{2} \hat{m}_{i j}F(j-i) \ell \right] + \exp\left[ \frac{1}{2} \hat{m}_{j i}F(i-j)\ell \right]}.
\label{bias-rate}
\end{eqnarray}
The argument within the exponential in the RHS is an extra energy cost for transferring mass $\hat{m}_{i j}=1$ from site $i$ to $j=i\pm 1$ and $\ell$ being the lattice spacing. Thus for small $F$, Eq.~\eqref{bias-rate} becomes $c^F_{i j} =c_{i j} \left[1+ F(j-i) \ell /2 \right]+{\cal O}(F^2)$ up to a leading order term in $F$. Incorporating the modified rates, the evolution of the local density is still a gradient type as given by \cite{supplement}
\begin{equation}
\frac{\partial \tilde{\rho}_i(t)}{\partial t} = \frac{1}{2} \left( G_{i-1}+G_{i+1}-2G_i \right) - \frac{F \ell}{4} \left( G_{i+1}-G_{i-1} \right).
\label{gradient}
\end{equation}
Now, going to the diffusive scaling limit by rescaling the lattice spacing $\ell=1 \rightarrow 1/\tilde{L}$, we obtain the hydrodynamic equation for the typical coarse-grained trajectories in the biased system as \cite{supplement}:
\begin{equation}
\frac{\partial \tilde{\rho}(x,\tau)}{\partial \tau} =  - \frac{\partial}{\partial x} \left( -D(\tilde{\rho}) \frac{\partial \tilde{\rho}}{\partial x} + \chi(\tilde{\rho})F \right),
\label{drift-diffusion}
\end{equation}
with the two transport coefficients
\begin{subequations}
\begin{equation}
D(\tilde{\rho})=\frac{1}{2}\frac{\partial G(\tilde{\rho})}{\partial \tilde{\rho}} ~~~~~\mbox{and}~~~~ \chi(\tilde{\rho})=\frac{1}{2}G(\tilde{\rho}),
\label{transport-unbounded}
\end{equation}
\begin{equation}
\mbox{with}~~~ G(\tilde{\rho})=qA(\tilde{\rho})+ pU(\tilde{\rho})/N_p
\label{observable-unbounded}
\end{equation}
\label{transport-unbounded1}
\end{subequations}
being an observable which is also a function of $p$, $T_0$, $N_p$ and $l_0$.

So far we have derived the hydrodynamics and calculated the transport coefficients for the UMTM model as a function of mass density $\tilde{\rho}$ whereas our original system is EM with particle density $\rho$. We use the reverse mapping of the MFT hydrodynamics \cite{Rizkallah_JSM2023} and recover the fluctuating hydrodynamics description for original EM with a similar structure as in Eq.~\eqref{fluc-hydro} with the two modified transport coefficients 
\begin{equation}
D({\rho})=-\frac{1}{2}\frac{\partial G({\rho})}{\partial {\rho}} ~~~~~\mbox{and}~~~~ \chi({\rho})=\frac{1}{2}\rho G({\rho}),
\label{transport-EP}
\end{equation}
with observable $G({\rho})$ of the same form as Eq.~\eqref{observable-unbounded} but replacing $\tilde{\rho}$ with $\rho$. Now in EM, the interpretation of $A(\rho)$ is the probability that the right neighboring site of a particle is empty and of $U(\rho)$ is the mean number of bound pili-pair on the right side provided that the right neighboring site is empty. Since the formation of bound pairs depends on the parameters in an involved way, it is difficult to obtain them explicitly. This also makes obtaining analytically the steady state particle number distribution challenging. However, one can bypass this difficulty by calculating the observables  $A(\rho)$ and $U(\rho)$ numerically and thus by getting the transport coefficients semi-analytically as a function of density, cell-cell interaction strength and other parameters. We perform this calculation and verify our semi-analytic results directly with microscopic simulation in the next section.

\begin{figure}
\begin{center}
\includegraphics[width=4.27cm,angle=0]{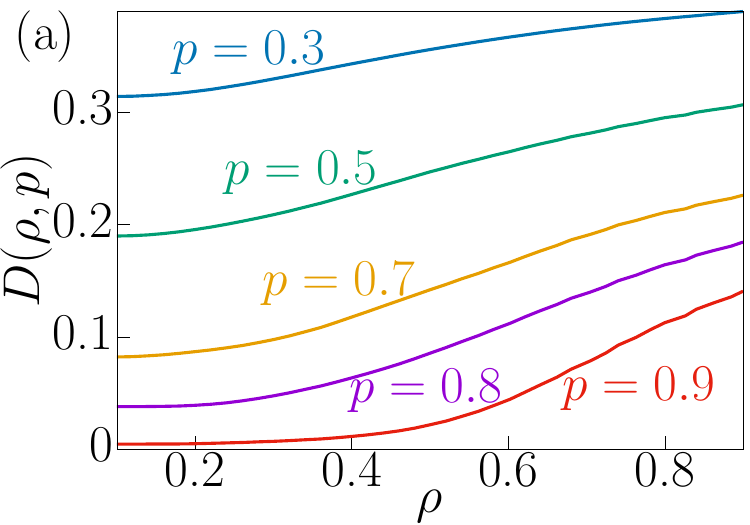}
\includegraphics[width=4.27cm,angle=0]{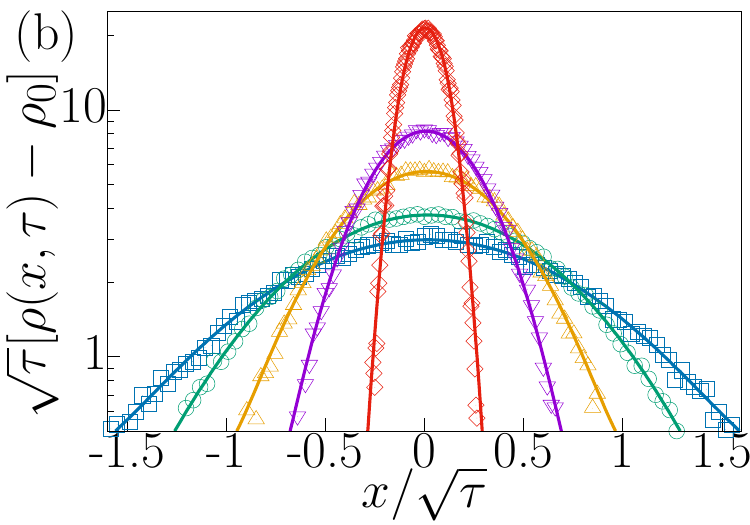}
\includegraphics[width=4.27cm,angle=0]{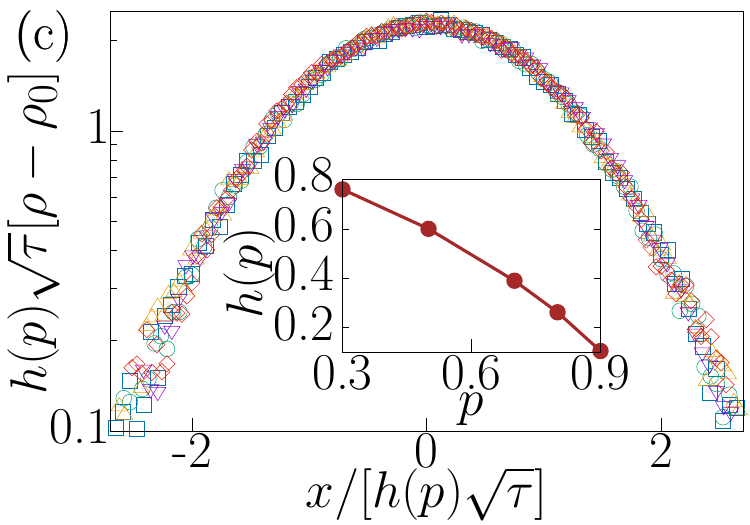}
\includegraphics[width=4.27cm,angle=0]{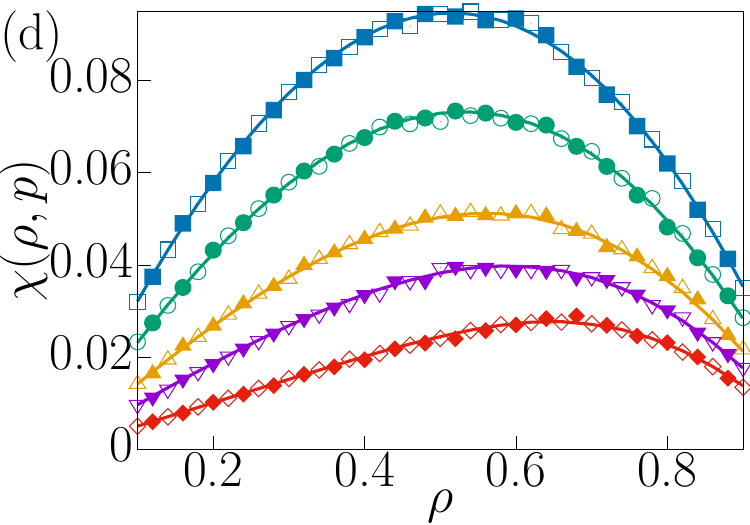}
\caption{Panel (a): Bulk-diffusivity $D(\rho,p)$ calculated from MFT is shown as a function of density $\rho$ for different values of the cell-cell interaction parameter $p$. $D(\rho,p)$ significantly decreases with increasing $p$ indicating a cluster formation. 
The color codes for legends in the following panels are the same as in panel (a). Panel (b): The MFT results of diffusivity is tested in simulation by looking at the scaled profile of initial narrow box-like density perturbation with highest density $\rho_h=0.875$ and with a background density $\rho_0=0.125$. Points are simulation results at a sufficiently large time $\tau$ whereas the lines are the numerical solution of diffusion equation with diffusivity given in panel (a). Panel (c) is presenting the scaling collapse of the density profiles with an empirical scaling factor $h(\rho)$ which is shown in the inset. Panel (d): Plot of conductivity $\chi(\rho,p)$ as a function of $\rho$ for different values of $p$. Lines are the MFT results and points are from simulations. Empty points are $\chi$ calculated from current fluctuation in an unbiased system and filled points are $\chi$ obtained from mean current in a biased system.}
\label{diffusivity}
\end{center}
\end{figure}

\section{Numerical results}
\label{sec:result}

\subsection{Transport coefficients}

We determine bulk-diffusivity $D(\rho)$ numerically using the formula \eqref{transport-EP} as a function of density $\rho$ and plot them for different activity of cell-cell interaction $p$ in panel (a) of Fig.~\ref{diffusivity} while we set other parameters at $T_0=10$, $N_P=10$, and $l_0=2$. In the equilibrium limit $p=0$, the bulk-diffusivity is constant at $D(\rho)=1/2$ \cite{Derrida_PRL2004, Chakraborty_PRE2020}. $D(\rho)$ is decreasing with increasing $p$ signifying a cluster formation at higher activity. We also observe, for a fixed $p$, a lower diffusivity at low density regime where the clustering is more prominent. We verify this behavior of bulk-diffusivity with simulations by comparing the spreading of a narrow boxlike density perturbation with the solution of the diffusion equation $\partial_\tau\rho=\partial_x[D(\rho)\partial_x \rho]$ starting from the same boxlike density profile where $D(\rho)$ is taken from the semi-analytic result in Eq.~\eqref{transport-EP}. In the panel (b) of Fig.~\ref{diffusivity}, represented by symbols, we show the rescaled density profiles calculated from microscopic simulations for different activities $p$. The lines are the (rescaled) solutions of the diffusion equation where the bulk-diffusivity $D(\rho)$ at different $p$ is taken from panel (a) of Fig.~\ref{diffusivity}. For each value of $p$, we checked density profiles from microscopic simulation at two different sufficiently large times. A remarkable agreement between lines and symbols establishes again the nominal diffusion transport in the clustered state. The density profiles can further be scaled to collapse as presented in panel (c) with a factor $h(p)$ which we found empirically and plot in the inset.

As mentioned earlier, one can notice from Fig.~\ref{diffusivity}(a) that for any fixed $p$, the bulk-diffusivity is lower at low densities. The key to understanding this phenomenon is in the distinction between the bulk diffusivity that we derive (related to diffusive flux) versus the single particle diffusivity (self-diffusion of a tagged particle) as mentioned above in Sec.~\ref{sec:hydro}.
In the equilibrium limit $(p=0)$, our model is equivalent to the simple symmetric exclusion process with density $\rho \in [0,1]$. In that case, the individual tagged particles move sub-diffusively with MSD $\sim t^{1/2}$. Therefore, the diffusion coefficient understood, in the classical sense, as the proportionality constant in linear dependence of the MSD as a function of time, can not be defined 
\cite{Alexander_PRB1978, Krapivsky_JSP2015, Krapivsky_PRL2014}. Yet the bulk-diffusivity based on the notion of diffusive flux is constant for all densities \cite{Derrida_PRL2004, Chakraborty_PRE2020}. As density approaches $\rho \sim 1$, there is a lack of vacancies in the system. This affects both movement of the particles but also the density gradient. Since it is difficult to create a large density gradient in such a high density, a small density gradient generates a proportionally small current (due to restricted movement of particles) keeping the bulk-diffusivity (defined as the ratio of the current to the density gradient) constant.

In our system, when $p>0$, clusters are created in the low density limit. This additional presence of the tight clusters limits the movement of particles but there is still space for creating significant density gradients. This keeps the bulk-diffusivity low. At higher density, the effect of clusters disappears as mean gap between the particles becomes shorter compared to the interaction length. As a result, on average, the number of bound pili-pairs for any particle is similar on both left and right sides. This lack of force imbalance results in the absence of clusters and enables particles to response to density gradients. One can also view this intuitively as a transition from localized isolated ``dense" clusters at overall low density of particles, to a more ``fluidized" continuum of particles at higher overall density.

Next, we numerically calculate conductivity $\chi(\rho)$ using the second relation in Eq.~\eqref{transport-EP}. In panel (d) of Fig.~\ref{diffusivity} we plot them with lines as a function of density $\rho$ for different cell-cell interaction parameter $p$ while the rest of the parameters are same as panel (a). The conductivity at equilibrium limit $p=0$ is known from the simple symmetric exclusion process as $\chi(\rho)=\rho (1-\rho)/2$ \cite{Derrida_PRL2004, Chakraborty_PRE2020}. Interestingly, the conductivity also decreases with an additional asymmetry arising across $\rho=0.5$ as we increase cell-cell interaction parameter $p$.  Direct computation of $\chi(\rho)$ from simulation also supports this finding -- which we perform in two ways. 
One way is by computing the fluctuation of the coarse-grained current $\sum_{i=1}^L j_i^{(t)}$ where $j_i^{(t)}$ is the time integrated current through bond $(i,i+1)$ during a long time $t$. Clearly, at steady state, for diffusive systems without a bias $\left \langle \sum_{i=1}^L j_i^{(t)} \right \rangle=0$, however, the fluctuation of this current accounting for the atypical coarse-grained trajectories is related to the conductivity through $\chi(\rho)=\left \langle \left( \sum_{i=1}^L j_i^{(t)} \right)^2 \right \rangle/(2Lt)$ \cite{Derrida_PRL2004, Tanmoy2023}. We show this conductivity in panel (d) of Fig.~\ref{diffusivity} by empty points which agree remarkably with the semi-analytic observations. Secondly, we apply a small bias $F=10^{-2}$ to the system following Eq.~\eqref{bias-rate} and numerically compute mean coarse-grained current $\left \langle j_i^{(t)} \right \rangle /t$. According to the EM version of Eq.~\eqref{drift-diffusion}, this current at steady state should be equal to $F \chi(\rho)$. The $\chi(\rho)$ calculated in this way is presented by solid points in panel (d) of Fig.~\ref{diffusivity} where the agreement with the same obtained semi-analytically is again excellent.

\begin{figure}
\begin{center}
    \includegraphics[width=8.5cm,angle=0]{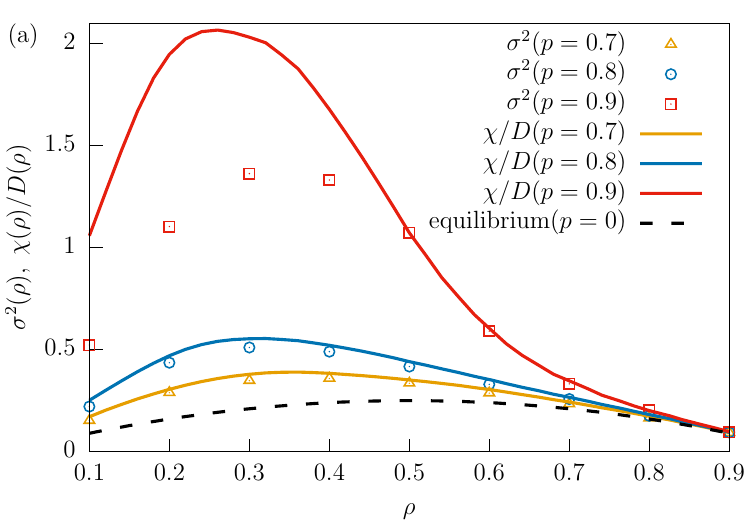}
\includegraphics[width=4.25cm,angle=0]{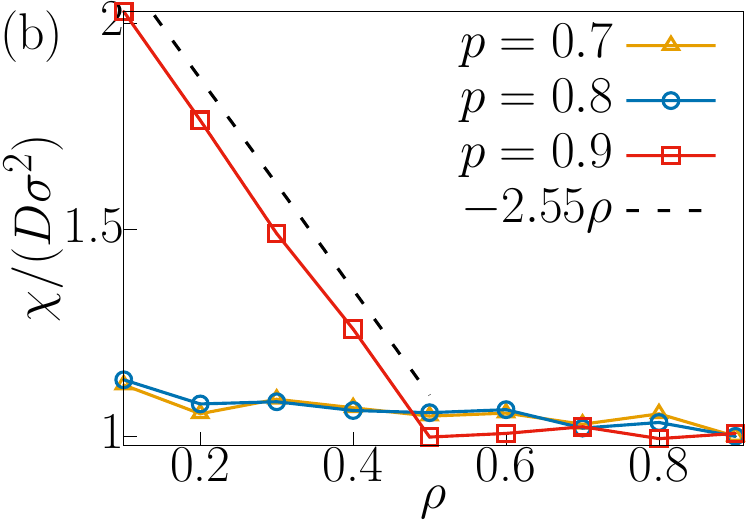}
\includegraphics[width=4.25cm,angle=0]{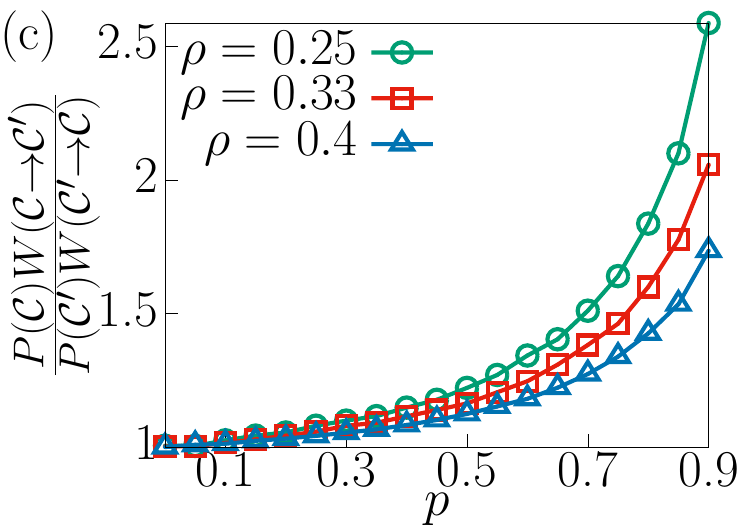}
\includegraphics[width=8.5cm,angle=0]{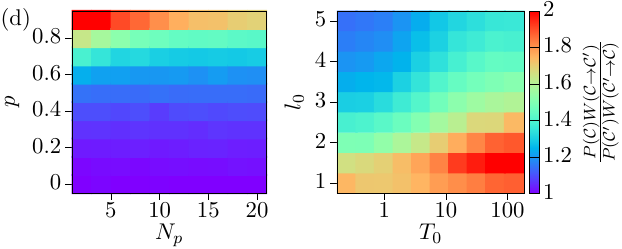}
\caption{Panel (a): Comparison of numerically obtained $\sigma^2(\rho)$ with the ratio of semi-analytically calculated transport coefficients $\chi(\rho)/D(\rho)$. $\sigma^2$ is computed numerically for system size $L=1000$ and subsystem size $s$ for which the $\sigma^2$ is maximum while we set other parameters $N_p=10$, $l_0=2$ and $T_0=10$ MCT. The black dashed line is $\chi(\rho)/D(\rho)=\sigma^2(\rho)=\rho (1-\rho)$ for equilibrium case with $p=0$. Panel (b): Plot showing the deviation of the ratio $\chi/\left(D \sigma^2 \right)$ from unity and thus establishing the breakdown of Einstein relation near the `clustering' limit at low $\rho$ and large $p$. Panel (c): Presentation of breakdown of microscopic detailed-balance while approaching the `clustering' limit. $N=2$ is fixed for all three plots with other parameters: $N_p=10$, $l_0=2 \ell$ and $T_0=10$ MCT. Panel (d): Dependence of the detailed-balance ratio on the key non-equilibrium parameters such as $p$, $N_p$, $T_0$ and $l_0$. For left panel we set $N=2$, $L=5$, $l_0=2 \ell$ and $T_0=10$ MCT. For the right panel we set $N=2$, $L=5$, $N_p=10$ and $p=0.9$.}
\label{detailed-balance}
\end{center}
\end{figure}

\subsection{Einstein relation}

To this end, we can also check one important consequence of MFT, the Einstein relation (ER), which in general holds for systems for which the trajectories of the macroscopic fields such as density and current are time reversible irrespective of whether the microscopic dynamics is equilibrium or not \cite{Bertini_PRL2001, Bertini_RMP2015}. With this condition of macroscopic reversibility, the probability large deviation function for those fields can be expressed as a quasi-potential from which one can derive the Einstein relation connecting the ratio of two transport coefficients to the scaled subsystem density fluctuation through $\chi(\rho)/D(\rho)=\sigma^2(\rho)$ where the scaled fluctuations of particle number $n$ within a subsystem of size $s$ is $\sigma^2(\rho)=\lim_{s \rightarrow \infty} \left[\left \langle n^2 \right \rangle - \left \langle n \right \rangle^2 \right]/s$ \cite{Bertini_PRL2001, Bertini_RMP2015, Derrida_JSM2007}. We compute $\sigma^2(\rho)$ numerically and compare them with the ratio $\chi(\rho)/D(\rho)$ in panel (a) of Fig.~\ref{detailed-balance} for different $p$. We observe that $\sigma^2(\rho)$ is always lower than $\chi(\rho)/D(\rho)$ and the deviation increases as we approach `clustering' limit at larger $p$ and lower $\rho$. To quantify this deviation we plot the ratio $\chi/(D\sigma^2)$ in panel (b) which should be unity if the Einstein relation is satisfied. The ratio deviates from unity in the clustering regime and it linearly depends on density $\rho$ with a negative slope, particularly when $\rho < 0.5$. We ensured to minimise the finite size effect while computing $\sigma^2(\rho)$ \cite{supplement}. This result is surprising because the UMTM version of our model has a structure of a special type of gradient system, known as the zero-range-process \cite{Evans_JPA2005} where the transition rates $c_{i,j}$ depend only on initial site $i$. For this class of systems the ER remained always valid in the past \cite{Bertini_RMP2015}. Therefore, the breakdown of ER in our model is non-trivial and it could be an outcome of the inherent out-of-equilibrium dynamics of our system which we can trace to a breakdown of detailed-balance at microscopic scale. 

\subsection{Broken detailed-balance}

Consider the simplest setting of $N=2$ particles in a system of size $L=5$. Then we denote a `clustered' configuration ${\cal C}:=\rule{0.25cm}{.5pt} ~\rule{0.25cm}{.5pt}~ \blacksquare~ \blacksquare~ \rule{0.25cm}{.5pt}$, and a `non-clustered' configuration ${\cal C}^\prime:=\rule{0.25cm}{.5pt} ~\blacksquare~\rule{0.25cm}{.5pt}~  \blacksquare~ \rule{0.25cm}{.5pt}$, where $\blacksquare$ and $\rule{0.25cm}{.5pt}$ respectively denote an occupied and an empty site. Provided that $P({\cal C})$ is the probability of finding the system in a configuration ${\cal C}$ at steady state and $W({\cal C}\rightarrow {\cal C}^\prime)$ is the transition rate from configuration ${\cal C}$ to ${\cal C}^\prime$, the microscopic condition of equilibrium \cite{Gardinerbook} is given by the detailed balance $\frac{P({\cal C})W({\cal C}\rightarrow {\cal C}^\prime)}{P({\cal C}^\prime)W({\cal C}^\prime\rightarrow {\cal C})} = 1$. Interestingly, the numerically computed $\frac{P({\cal C})W({\cal C}\rightarrow {\cal C}^\prime)}{P({\cal C}^\prime)W({\cal C}^\prime\rightarrow {\cal C})}$ significantly deviates from unity in the high activity regime as seen in panel (c) of Fig.~\ref{detailed-balance} leaving the only equilibrium limit at $p=0$. The deviation of this ratio from unity is more as we approach lower density by increasing blank sites to the right end of both ${\cal C}$ and ${\cal C}^\prime$. Thus we confirm that the clustering phenomena in the high $p$ and low $\rho$ regime is intrinsically non-equilibrium. In panel (d), we present the role of other key parameters in non-equilibrium clustering through a heat-map of $\frac{P({\cal C})W({\cal C}\rightarrow {\cal C}^\prime)}{P({\cal C}^\prime)W({\cal C}^\prime\rightarrow {\cal C})}$. The deviation from unity indicates that the system is more non-equilibrium with increasing $p$ and $T_0$ and with decreasing $N_p$ and $l_0$. The mean pili-lifetime $T_0$ plays a role of persistence time, however, its role is mostly notable at the cluster boundaries because within a cluster the number difference in left-right pili-pairs of a particle is insignificant as the mean gap is low. The argument is similar for mean pili-length $l_0$. At higher $l_0$ and at higher $N_p$ the scaled number difference in left-right pili-pair [$(n^L-n^R)/N_p$] of a particle  decreases resulting in an equilibrium-like unbiased movement.

\section{Conclusions}
\label{conclusion}

We provided a realistic but simple model of active cellular aggregates with pili-mediated interactions which shows non-equilibrium clustering in one dimension. Using a recently developed Macroscopic Fluctuation Theory, we formulated a fluctuating hydrodynamic description from which we could clearly identify two density dependent transport coefficients -- bulk diffusivity and conductivity. Both of these hydrodynamic transports can be obtained through two simple observables -- $A(\rho)$, the probability that the right (left) neighboring site of a particle is empty so that it can move and $U(\rho)$, the mean number of bound pili-pairs on the right (left) side provided that the right (left) neighboring site is empty so that it can be pulled. These two observables can be potentially assessed analytically or experimentally. Lack of knowledge of explicit analytical dependence of particle distribution on the non-equilibrium parameters makes further analytical progress challenging. Therefore, we computed the observables at steady state from simulations and thus avail the semi-analytic hydrodynamic transport coefficients. Both transport coefficients decrease with active cell-cell interaction parameter $p$ with breakdown of symmetry across $\rho=1/2$. The semi-analytic theory was then tested with direct simulations. The bulk diffusivity is verified by looking at the spreading of density perturbation with time according to Fick's law whereas the conductivity is tested in two ways. One way is from Ohm's law by computing mean current after applying an external biasing force. Another way is by evaluating macroscopic current fluctuations. Both the transport coefficients obtained from our semi-analytic theory are captured remarkably well from microscopic simulations. However, we observed numerically that the Einstein relation connecting the ratio of the transport coefficients to the subsystem density fluctuation does not hold in this model in the limit of high $p$ and low $\rho$. Correspondingly, the detailed balance -- the cornerstone of equilibrium also breaks down in the same limit. The detailed relation between these two observations is beyond the scope of this paper and could be investigated in future.

A crucial ingredient to satisfy ER is the \textit{macroscopic reversibility} \cite{Bertini_RMP2015}. Therefore, examining the macroscopic irreversibility could be an interesting question to investigate the breakdown of ER for this system. Another immediate outlook 
is accounting for the cell-substrate interactions in more detail. 
Of course an open and outstanding problem that remains is obtaining an exact analytical expression for the transport coefficients. It would be important to extend the framework to derive fluctuating hydrodynamics for other biological systems of multicellular aggregates such as tumor spheroids (microscopically interacts through cell adhesion and contraction driven tensile forces) \cite{Wolfgang_Review1987, Friedl_NatureReview2003, Fabry_PLoS2012, Fabry_eLife2020} or neuronal organoids (with microscopic activity of migration, dissociation and aggregation) \cite{Karow_Nature2018, Karow_2021, Lancaster_Nature2013, Pasca_Nature2018}. 

The transport coefficients can be obtained experimentally also. For both transport coefficients the common way to calculate them is by measuring cell-currents. To compute cell current experimentally in a 2D system, one could consider a small area of size $l\times l$ of the entire colony and count $n_c$ -- the total number of cells passing through that subsection in a particular direction (say, along positive $x$-axis) within a period of time $t$. Then $J_c=n_c/(lt)$ is the local cell current in the system. The bulk-diffusivity can be determined from the mean cell-current measured in presence of a density gradient in the same direction. Another practical way to obtain $D(\rho)$ is from the time dependent amplitude of a sinusoidal density perturbation as described in \cite{Tanmoy_PRE2024}. The proportionality coefficient of the exponent of the time dependent amplitude gives the bulk-diffusivity. 
Similarly, the conductivity can be accessed by measuring the cell-current after applying a small external bias. A relevant external force in a biological system could be a gravitational force when a substrate on which cells move is inclined relative to the horizontal orientation. If a larger force is needed, centrifuging the colonies might be an option. For samples that are studied immersed in solution, the application of the shear flow could be used as means of generating a laterally uniform directional force.

While the proposed theory provides basis for qualitative understanding the fluctuating hydrodynamics in an idealized scenario of a one-dimensional system, it would be important to push this theory to two and three-dimensional settings as observed in the respective experimental situations but where the mapping to UMTM is not applicable.  It would be also exciting to understand the connection of the transport coefficients obtained by the theory to the material properties of the cellular aggregates.

\section*{ACKNOWLEDGMENTS}

We thank Punyabrata Pradhan for helpful discussions and comments on the draft of the manuscript. VZ acknowledges financial support by the German Research Foundation (DFG) project 460333672 CRC1540 EBM.

\begin{widetext}

\subsection*{Appendix A: Derivation of hydrodynamics}

The time update equation for the mass $m_i(t)$
\begin{eqnarray}
m_i(t+dt) =
\left\{
\begin{array}{ll}
m_i(t) - 1            & {\rm prob.}~ c_{i,i-1} dt/2, \\
m_i(t) - 1            & {\rm prob.}~ c_{i,i+1} dt/2, \\
m_i(t) + 1            & {\rm prob.}~ c_{i-1,i} dt/2, \\
m_i(t) + 1            & {\rm prob.}~ c_{i+1,i} dt/2, \\
m_i(t)                & {\rm prob.}~ 1-\Sigma dt,
\end{array}
\right.
\label{unbounded-unbiased-s}
\end{eqnarray}
with $\Sigma=\left(c_{i,i-1}+c_{i,i+1}+c_{i-1,i}+c_{i+1,i}\right)/2$ and the jump weights $c_{i,i-1}=qa_i+pa_in^R_i/N_p$, $c_{i,i+1}=qa_i+pa_in^L_{i+1}/N_p$, $c_{i-1,i}=qa_{i-1}+pa_{i-1}n^L_i/N_p$ and $c_{i+1,i}=qa_{i+1}+pa_{i+1}n^R_{i+1}/N_p$ with $a_i=1-\delta_{m_i,0}$ is the occupation probability site $i$. After implementing the condition $n^R_i=n^L_{i+1}$ for any $i$ we get $c_{i,j}=qa_i+pa_in^R_i/N_p$ depends only on the initial site $i$. In addition to that, we apply a small bias $F$ in the right direction and thus the jump weights will be modified as $c^F_{i j} =c_{i j} \left[1+ F(j-i) \ell /2 \right]+{\cal O}(F^2)$ up to leading order term in $F$. Incorporating this into Eq.~\eqref{unbounded-unbiased-s} we reach the mass update equation for a biased system as:
\begin{eqnarray}
m_i(t+dt) =
\left\{
\begin{array}{ll}
m_i(t) - 1            & {\rm prob.}~ (qa_i+pa_in^R_i/N_p) (1+ F \ell /2) dt/2, \\
m_i(t) - 1            & {\rm prob.}~ (qa_i+pa_in^R_i/N_p) (1- F \ell /2) dt/2, \\
m_i(t) + 1            & {\rm prob.}~ (qa_{i-1}+pa_{i-1}n^R_{i-1}/N_p) (1+ F \ell /2) dt/2, \\
m_i(t) + 1            & {\rm prob.}~ (qa_{i+1}+pa_{i+1}n^R_{i+1}/N_p) (1- F \ell /2) dt/2, \\
m_i(t)                & {\rm prob.}~ 1-\Sigma_F dt,
\end{array}
\right.
\label{unbounded-biased-s}
\end{eqnarray}
Now we show the master equation structure for the local mass density $\tilde{\rho}_i(t)=\langle m_i(t) \rangle$ as:
\begin{eqnarray}
[\tilde{\rho}_i(t+dt)-\tilde{\rho}_i(t)]/dt = (\tilde{\rho}_i-1)G_i + (\tilde{\rho}_i+1)[G_{i-1}(1+ F \ell /2)+G_{i+1}(1- F \ell /2)]/2 - \tilde{\rho}_i \Sigma_F ,
\end{eqnarray}
with $G_i=q A_i + \frac{p}{N_p}U_i$. After simplification, this gives
\begin{equation}
\frac{\partial \tilde{\rho}_i(t)}{\partial t} = \frac{1}{2} \left( G_{i-1}+G_{i+1}-2G_i \right) - \frac{F \ell}{4} \left( G_{i+1}-G_{i-1} \right).
\label{gradient-s}
\end{equation}
Now, going to the diffusive scaling limit by rescaling space $i \rightarrow x = i/\tilde{L}$, time $t \rightarrow \tau = t/\tilde{L}^2$ and the lattice spacing  $\ell=1 \rightarrow 1/\tilde{L}$, we obtain the hydrodynamic equation in the biased system as:
\begin{eqnarray} 
\frac{\partial \tilde{\rho}(x,\tau)}{\tilde{L}^2 \partial \tau} = \frac{1}{2} \left[ G \left( x-\frac{1}{\tilde{L}}, \tau \right) + G \left( x + \frac{1}{\tilde{L}}, \tau \right) - 2G(x, \tau) \right] - \frac{1}{4} \left[ G \left( x+\frac{1}{\tilde{L}}, \tau \right) - G\left( x - \frac{1}{\tilde{L}}, \tau \right) \right] \frac{F}{\tilde{L}}.
\label{grad-exp1}
\end{eqnarray}
Next we expand the function $G$ around space $x$ for small ${\cal O}(1/L)$ as following:
\bea
 G \left( x \pm \frac{1}{\tilde{L}}, \tau \right) = G(\rho(x, \tau)) \pm \frac{1}{\tilde{L}} \frac{\partial G(\rho(x,\tau))}{\partial x} + \frac{1}{2 \tilde{L}^2} \frac{\partial^2 G(\rho(x, \tau))}{\partial x^2} + {\cal O} \left( \frac{1}{\tilde{L}^3} \right).
\eea
After substituting this expansion into Eq.~\eqref{grad-exp1} and keeping terms up to order $1/\tilde{L}^2$ we obtain the hydrodynamic equation for the biased system as
\bea
\frac{\partial \tilde{\rho}(x,\tau)}{\tilde{L}^2 \partial \tau} = \frac{1}{2} \left[ \frac{1}{\tilde{L}^2} \frac{\partial ^2 G(\tilde{\rho})}{\partial x^2} \right] - \frac{1}{4} \left[ \frac{2}{\tilde{L}} \frac{\partial G(\tilde{\rho})}{\partial x} \right] \frac{F}{\tilde{L}} +{\cal O} \left( \frac{1}{\tilde{L}^3} \right).
\eea
\end{widetext}

Therefore,
\bea
\nn
\frac{\partial \tilde{\rho}(x,\tau)}{\partial \tau} &=& - \frac{\partial}{\partial x} \left(-\frac{1}{2}\frac{\partial G(\tilde{\rho})}{\partial \tilde{\rho}} \frac{\partial \tilde{\rho}}{\partial x} + \frac{1}{2}G(\tilde{\rho}) F  \right) \\
&=&  - \frac{\partial}{\partial x} \left( -D(\tilde{\rho}) \frac{\partial \tilde{\rho}}{\partial x} + \chi(\tilde{\rho})F \right),
\label{drift-diffusion-s}
\eea
with the two transport coefficients
\begin{equation}
D(\tilde{\rho})=\frac{1}{2}\frac{\partial G(\tilde{\rho})}{\partial \tilde{\rho}} ~~~~~\mbox{and}~~~~ \chi(\tilde{\rho})=\frac{1}{2}G(\tilde{\rho}).
\label{transport-unbounded-s}
\end{equation}

\subsection*{Appendix B: Role of other parameters on conductivity}

\begin{figure}
\begin{center}
\includegraphics[width=4.27cm,angle=0]{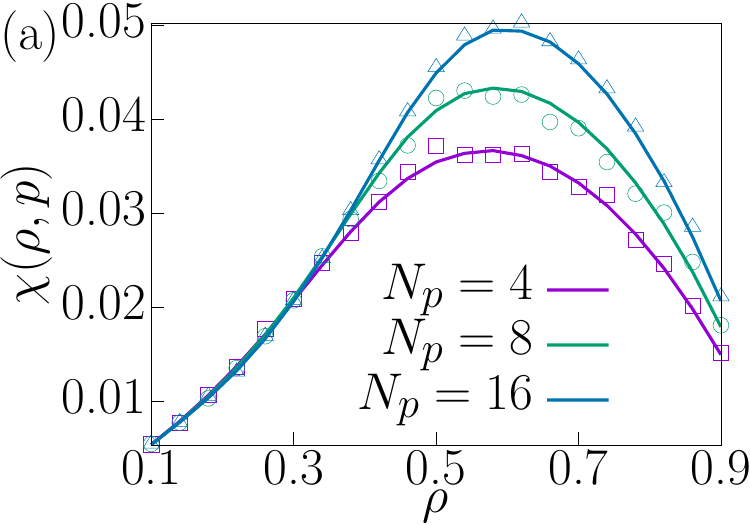}
\includegraphics[width=4.27cm,angle=0]{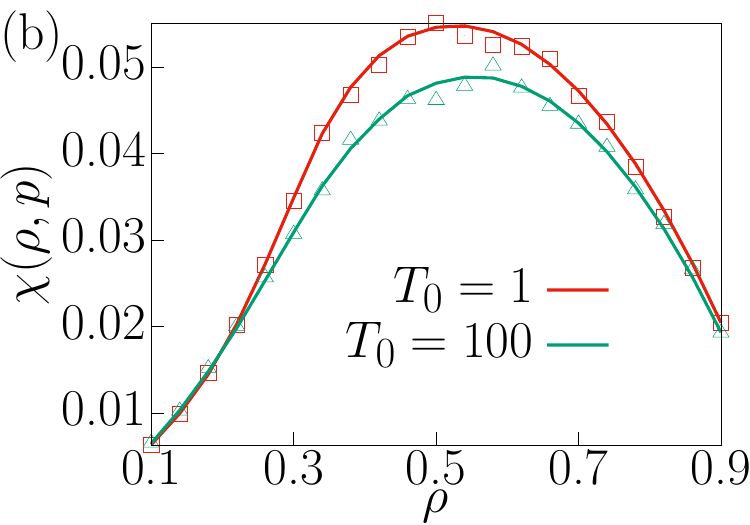}
\caption{Plot of conductivity $\chi(\rho)$ for different non-equilibrium parameters. Panel (a): Dependence on $N_p$ while keeping other parameters fixed at $p=0.9$, $T_0=10$ and $l_0=5$. Panel (b): Dependence on $T_0$ while keeping other parameters fixed at $p=0.9$, $N_p=10$ and $l_0=10$.}
\label{conductivity-s}
\end{center}
\end{figure}

We now show the behavior of the conductivity $\chi(\rho)$ as a function of the two other non-equilibrium parameters pili-number per particle $N_p$ and mean pili lifetime $T_0$. In panel (a) of Fig.~\ref{conductivity-s} we show the dependence on $N_p$ while setting other parameters at $p=0.9$, $T_0=10$ and $l_0=5$. The dependence on $T_0$ is presented in panel (b) of Fig.~\ref{conductivity-s} keeping fixed $p=0.9$, $N_p=10$ and $l_0=10$. In both panels, the liens are observed by semi-analytic calculations and the points are the conductivity numerically computed for unbiased system using the formula $\chi(\rho)=\left \langle \left( \sum_{i=1}^L j_i^{(t)} \right)^2 \right \rangle/(2Lt)$ \cite{Derrida_PRL2004, Tanmoy2023}. In both cases, the conductivity is affected mostly in the intermediate density regime.

\begin{figure}
    \centering
    \includegraphics[width=4.27cm,angle=0]{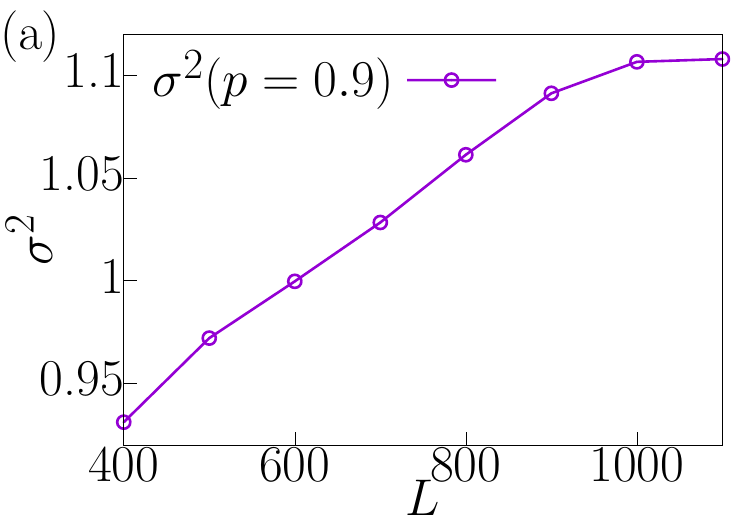}
    \includegraphics[width=4.27cm,angle=0]{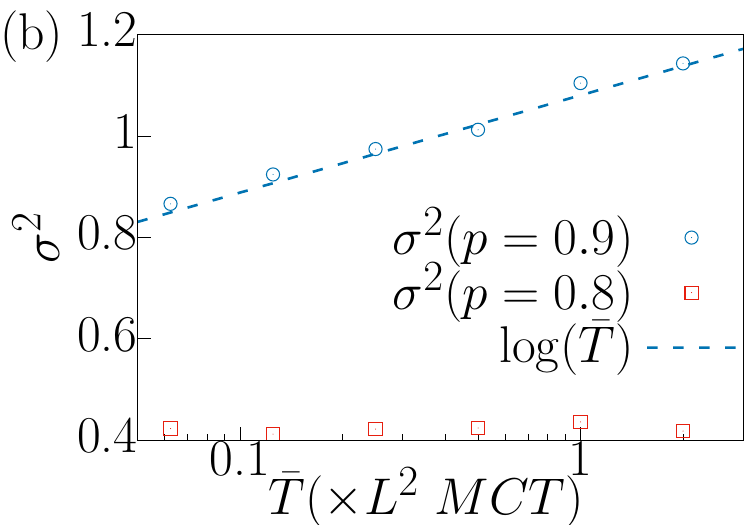}
    \caption{Panel (a): Testing the finite size effect by plotting $\sigma^2(\rho)$ at $p=0.9$ and $\rho = 0.2$ for different system sizes $L$ with other parameters $N_p=10$, $l_0=2$ and $T_0=10$ MCT. $\sigma^2(\rho)$ saturates as we increase system size $L$ leading to minimisation of finite size effect. Panel (b): We point out the role of the time $\bar{T}$ we leave to reach the system steady state. In the extreme clustering regime with $p=0.9$ and $\rho=0.2$, the $\sigma^2$ grows slowly with $\log(\bar{T})$ whereas in the light clustering regime with $p=0.8$ and $\rho=0.2$, $\sigma^2$ rapidly saturates with $\bar{T}$.}
    \label{finite-size}
\end{figure}

\subsection*{Appendix C: Effect of finite size and steady-state time on subsystem density fluctuation}

In this section, we demonstrate that our numerical results while computing the scaled fluctuations $\sigma^2(\rho)=\lim_{s \rightarrow \infty} \left[\left \langle n^2 \right \rangle - \left \langle n \right \rangle^2\right]/s$ of particle number $n$ within a subsystem of size $s$ have the minimum finite size effect. This definition of the scaled particle number fluctuations is valid when the subsystem size $s$ and the system size $L$ both are large with the condition $s\ll L$ such that the system can act as a particle reservoir to the subsystem. In Fig.~\ref{finite-size}(a), we show $\sigma^2(\rho)$ for different system sizes $L$ at an extreme non-equilibrium regime of $p=0.9$ and $\rho=0.2$. We choose the $s$ such that the $\sigma^2(\rho)$ is maximum for a given $L$. We observe that $\sigma^2(\rho)$ converges with system size $L$. However, as seen in panel (b) of Fig.~\ref{finite-size}, the time $\bar{T}$ we leave to reach the system steady state while computing the $\sigma^2(\rho)$ plays an important role. For system size $L=1000$, in the lightly clustering regime ($\rho=0.2$ and $p=0.8$) the fluctuation $\sigma^2$ saturates rapidly with $\bar{T}$, whereas in the extreme clustering regime ($\rho=0.2$ and $p=0.9$), $\sigma^2$ grows with $\log(\bar{T})$ indicating a slow growth of cluster size. This slow growth of clusters has been observed before \cite{Zaburdaev2015, Zaburdaev_PRE2015} in the system of {\it N. gonorrhoeae} bacterial colony formation. This phenomenon could be responsible for the \textit{macroscopic irreversibility} mentioned in Ref. \cite{Bertini_RMP2015} and thus might be the key to the breakdown of Einstein relation in the extreme non-equilibrium limit. Moreover, waiting for an excessively long period of time is not realistic in the context of the original biological problem where the cellular aggregates are observed for a finite, and typically not too long periods of time.

\subsection*{Appendix D: Pili-bound clustering}

\begin{figure}
\begin{center}
\includegraphics[width=4.27cm,angle=0]{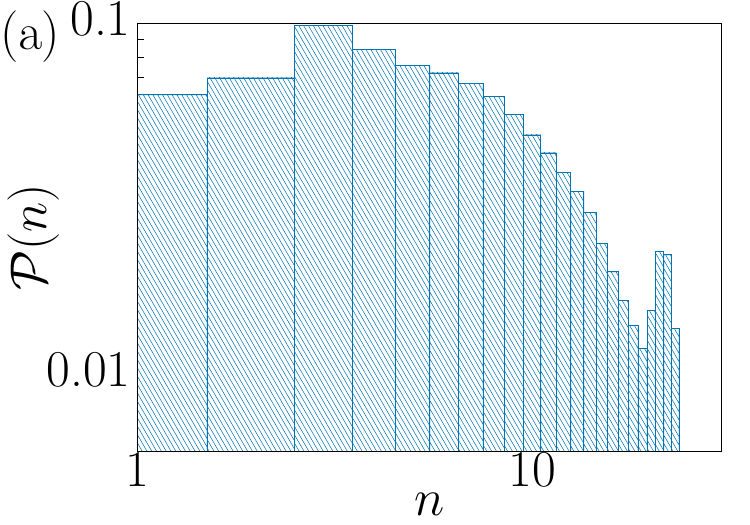}
\includegraphics[width=4.27cm,angle=0]{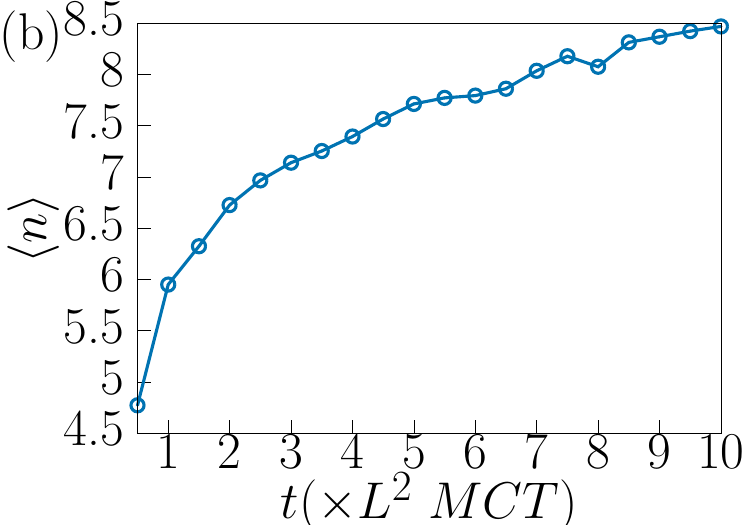}
\caption{Panel (a): Probability distribution ${\cal P}(n)$ of cluster size $n$ for $N=20$ particles in a system of size $L=100$ at $p=0.9$. Panel (b) shows the growth of mean cluster size $\langle n \rangle$ with time $t$ indicating a coalescence of clusters.}
\label{cluster-s}
\end{center}
\end{figure}

 In Figs.~\ref{trajectories}(b-c) a cluster is defined as a sequence of consecutive sites occupied by particles without a gap between them. We now redefine a cluster as a sequence of particles connected by interactions of bound pili-pairs. Consequently, any gap between two particles connected by bound pairs is considered part of the cluster.
We plot the new cluster size distribution in Fig.~\ref{cluster-s}(a) where we observe the appearance of a ``hump". This is similar to the distribution that was previously reported in experiments of clustering \textit{N. gonorrhoeae} bacteria in Ref.~\cite{Zaburdaev2015}. In panel (b) we can also see how the mean cluster size grows in time as observed in Ref.~\cite{Zaburdaev2015}, indicating the coalescence of smaller clusters into larger ones.

\bibliography{proposal}

\end{document}